# AN OPTIMAL BASELINE SELECTION METHODOLOGY FOR DATA-DRIVEN DAMAGE DETECTION AND TEMPERATURE COMPENSATION IN ACOUSTO-ULTRASONICS

M.-A. Torres-Arredondo[1], Julián Sierra-Pérez[2,*], Guénaël Cabanes[3]

[1] *MAN Diesel & Turbo SE, Engineering 4 Stroke, Mechanics, Measurement (EEDTTM), Augsburg, Germany*
[2] *Grupo de Investigación en Ingeniería Aeroespacial GIIA; Aeronautics Engineering School; Universidad Pontificia Bolivariana, Sede Medellín, Circular 1 70-01, Medellín, Colombia*
[3] *Université Paris 13, Institut Galilée, LIPN - CNRS, France*

*Email: julian.sierra@upb.edu.co

**ABSTRACT**

The global trends in the construction of modern structures require the integration of sensors together with data recording and analysis modules so that their integrity can be continuously monitored for safe-life, economic and ecological reasons. This process of measuring and analyzing the data from a distributed sensor network all over a structural system in order to quantify its condition is known as structural health monitoring (SHM). Guided ultrasonic wave-based techniques are increasingly being adapted and used in several SHM systems which benefit from built-in transduction, large inspection ranges, and high sensitivity to small flaws. Nonetheless, for the design of a trustworthy health monitoring system, a vast amount of information regarding the inherent physical characteristics of the sources and their propagation and interaction across the structure is crucial. Moreover, any SHM system which is expected to transition to field operation must take into account the influence of environmental and operational changes which cause modifications in the stiffness and damping of the structure and consequently modify its dynamic behaviour. On that account, special attention is paid in this paper to the development of an efficient SHM methodology where robust signal processing and pattern recognition techniques are integrated for the correct interpretation of complex ultrasonic waves within the context of damage detection and identification. The methodology is based on an acousto-ultrasonics technique where the discrete wavelet transform is evaluated for feature extraction and selection, linear principal component analysis for data-driven modelling and self-organizing maps for a two-level clustering under the principle of local density. At the end, the methodology is experimentally demonstrated and results show that all the damages were detectable and identifiable.

**KEYWORDS :** *Damage Detection, Acousto-Ultrasonics, Signal Processing, Pattern Recognition, Wavelet Transform, Principal Component Analysis, Self-Organizing Maps, Temperature Compensation.*

**INTRODUCTION**

During the last decades, guided waves have shown great potential for Structural Health Monitoring (SHM) applications. These waves can be excited and sensed by piezoelectric elements that can be permanently attached onto a structure offering online monitoring capability [1-3]. The advantages given by using guided ultrasonic waves have motivated an enormous amount of research using Lamb waves for the recognition and investigation of critical flaws in structures. Nevertheless, in reality, structures are susceptible to varying environmental and operational conditions (EOC) which affect the measured signals [4]. Studies have shown that temperature and damage can have generate a comparable impact on the dynamic behaviour of a structure [5]. These environmental and operational changes of the system can often mask slight changes in the structural dynamic responses caused by damage and covert the damage detection step into a very complex process [6]. For this reason, it is very important to take these conditions into account so that a reliable statement about the health of the structure under changing environmental and operational conditions with a minimum of false alarms can be accomplished.

For example, Fritzen et al. [7] modified an existing subspace-based identification method with temperature compensation for damage diagnosis. Moll et al. [8] studied the compensation of environmental influences using a simulation model and a laboratory structure. Buethe et al. [9] evaluated self-organizing maps in order to distinguish between environmental changes and damage of within a structure so that false alarm minimization could be accomplished under changing environmental conditions. Torres and Fritzen [10] presented theoretical developments, numerical and experimental results in order to analyze the effects of all the aforementioned sources of variability on wave propagation velocities, directionality and attenuation. Lu and Michaels [11] performed studies in order to find selective features which are sensitive to damage but insensitive to the applied surface wetting. Croxford et al. [12] evaluated two different methods to compensate for the temperature effect, namely optimal baseline selection (OBS) and baseline signal stretch (BSS). Kraemer et al. [13] proposed an approach based on Artificial Neural Networks (ANN) using Self Organizing Maps (SOM) in order to compensate the temperature effects on different features obtained from measured time data of a Carbon Fibre Reinforced Polymer plate (CFRP).

A number of authors have addressed the damage identification task employing either approaches based on physical models [14] or statistical models generated from recorded measurements [15]. As Farrar and Worden proposed, the problem of damage detection can be solved following statistical pattern recognition [16]. Within this concept, data analysis plays one of the most important aspects for pattern recognition so that a reliable health monitoring strategy can be defined [17]. Within the context of this paper, a combined pattern recognition methodology together with feature extraction and multivariate analysis is proposed for the detection of damage from the analysis of recorded structural dynamic responses avoiding the implementation of a complex physical model. Normally, active strategies for structural health monitoring using ultrasonic guided waves mainly deal with excitation signals that are band limited in order to minimize the effect of dispersion [18]. However, in this work an acousto-ultrasonics approach is evaluated. The acousto-ultrasonics (AU) technique was originally developed in the late seventies as a non-destructive tool for the evaluation of the mechanical properties of composite materials [19]. Since then the AU technique has been used in order to assess and quantify damage in composite materials. This technique works in a frequency range similar to that used in acoustic emission. A review examining this technology and discussing several applications and monitoring scenarios in aeronautic and aerospace structures can be found in [20-21]. The acousto-ultrasonic approach presented in this paper is based on collecting all the waveform energy that is available, i.e. instead of selecting specific wave packets from the recorded signal, all the multiple reverberations are collected for their subsequent analysis [22].

This contribution extends and finalizes previous works from the authors in the field of damage detection and classification [23-24]. Nevertheless, the structure of this article is to be self-contained. Within this context, pattern recognition is presented and evaluated as a tool for optimal baseline selection in which baseline signals recorded at different temperature conditions are clustered in different groups and then used as optimal baseline models for a multi-variate statistical model based on principal component analysis (PCA). This step guarantees not only a reduction of the total size of the model to be employed, since not all the recorded data are required for building the models, but also provides a reduction in the variability inside the models so that a better description of the process which is desired to be described can be achieved. On that account, this paper is concerned with the experimental validation of a data-driven structural health monitoring methodology where a damage detection and classification scheme based on an acousto-ultrasonic (AU) approach under changing temperature conditions is followed. The layout of this paper is as follows. In the first part, an analysis of the effects of variable temperature and operation conditions on wave propagation is shown. The second part introduces the required background for understanding the proposed algorithms and methodology for damage detection. The third part presents the analysis of the different performed experiments together with the discussion of the results. Finally, concluding remarks are given in the last part.

**GUIDED WAVE SENSITIVITY ANALYSIS**

It is well-known that temperature variations may change the material properties of a structure [25]. Also factors such as moisture content and material age affect significantly the wave propagation characteristics in the material. In order to experimentally show the changes introduced by these effects, a simple composite plate made of 4 equal layers with a total thickness of 1.7*mm* and stacking of [0° 90° 90° 0°] was analyzed. For the experiments the temperature was raised stepwise up to $T=60 \pm 2°C$ in a temperature-controlled oven. The temperature was recorded using two PT100 sensor attached on the opposite corners of the structure.

Figure 1(a) shows the structure with dimensions 200*mm*×250*mm*. Nine piezoelectric transducers PIC151 were installed on the surface of the plate. The excitation voltage signal (at $P_5$) was a 12*V* Hann-windowed toneburst with a carrier frequency of 30*kHz* with 5 cycles. Figure 1(b) shows that the shape of the signal changes significantly due to the wet surface influence. Without any loss of generality, and due to the symmetry in wave propagation of the composite plate, only the signal behaviour from sensors two, three and six are discussed. It can be observed how the signal responses for sensor two and three are reduced with respect to their peak-to-peak magnitude with increasing temperature (Figure 1(c)). The inverse effect can interestingly be seen with sensor six. The observed time-shifts are caused by both thermal expansion and changes in wave velocities with temperature. The attenuation of Lamb wave can be linked to ultrasonic wave dispersion as a result of frequency dependent phase velocities and attenuation loss due to frequency/temperature dependent material damping. In this analysis, the effects caused by temperature on the transducer performance have not been study. However, these effects are significantly less than the effect of temperature on wave propagation within the structure [26]. However, the capacitance of piezoelectric (PZT) materials is temperature sensitive and increases as temperature rises what modifies the sensor response [27].

Other effects than temperature variations may also change the material properties of a structure. Factors such as material age effect, moisture content and structure operation affect significantly the wave propagation characteristics in the material. For example, it was observed that only 1000h of cyclic exposure to both ultra-violet (UV) radiation and condensation resulted in a 29% decrease in the transverse tensile strength of a carbon fibre-reinforced epoxy material [28]. Other studies have shown tensile strength reduction of about 40% due to moisture and temperature changes [29-30]. It has also been studied the deterioration of composite material properties with time when the composites are subject to freeze-thaw cycles, where a 25% reduction in ultimate strain capacity was found [31]. For this reason, and for the sake of clarity, a second example is presented in order to depict the changes in group velocities in relation to the material properties within a glass fibre reinforced plastic (GFRP) plate only for the fundamental guided-wave modes. More details about wave propagation in solid media and the evaluated plate can be found in [32-33]. Figure two shows the group velocities for reduction of 20% for each material constant at 60*kHz*. It can be seen that both variations of $E_{11}$ and $E_{22}$ have a strong influence on the velocities for the $S_0$ mode. These influences are reflected for the $SH_0$ mode only on its caustics and are practically not visible for the $A_0$ mode. The shear modulus $G_{12}$ has a small effect on the group velocities of the $S_0$ mode at +/-45° (and mirrored angles) and almost no influence at 0°/90°. The $A_0$ mode is nearly not affected. However, the effect of changing $G_{12}$ is quite strong for the $SH_0$ mode. The shear moduli $G_{13}$ and $G_{23}$ have a noticeable influence on the $A_0$ mode and almost no influence for the $S_0$ and $SH_0$ modes.

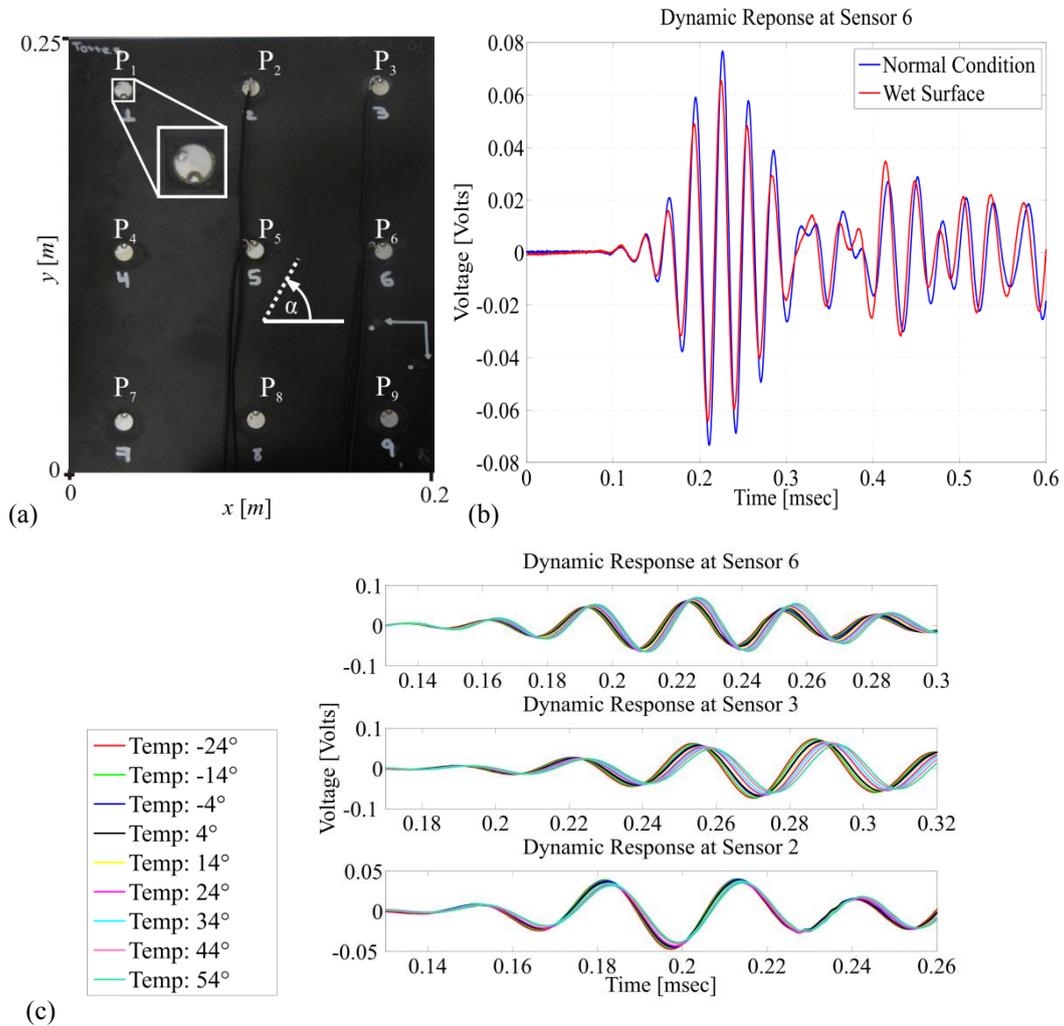

Figure 1. EOC Effects: (a) Studied Composite Plate, (b) Wet Surface Effect on Signal Amplitude, (c) Temperature Effects on Ultrasonic Wave Directionality.

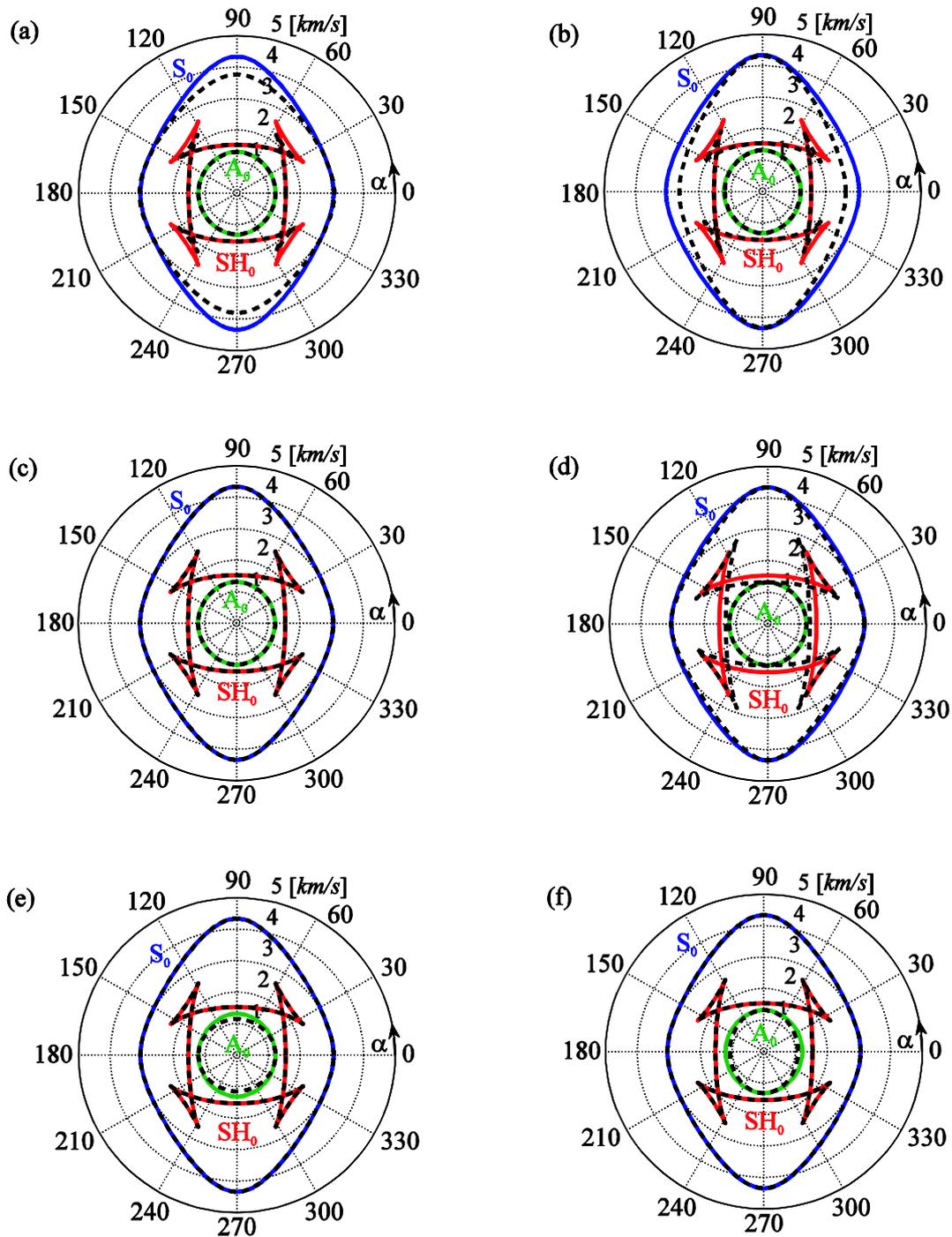

Figure 2. Effects of the material properties changes on ultrasonic wave curves for a reduction of 20% in: (a) $E_{11}$, (b) $E_{22}$, (c) $E_{33}$, (d) $G_{12}$, (e) $G_{13}$ and (f) $G_{23}$. Solid lines represent the results for the original values and the dashed lines the results for the reduced values.

**DIGITAL SIGNAL PROCESSING TECHNIQUES FOR DAMAGE DETECTION**

We propose in this paper a new method for damage detection based on a feature extraction using the Discrete Wavelet Transform, a Principal Component modelling and a clustering process computed with a density-based Self-Organising Map.

A brief introduction to the theoretical background for the different algorithms evaluated in the present paper is presented here. For a more detailed description regarding the mathematical background and applications please see [34-36].

*Feature Selection and Extraction by Means of the Discrete Wavelet Transform*

The discrete wavelet transform (DWT) is scientifically used for the analysis of signals where their complex changes are represented by transformation coefficients at a specified time and scale [37]. The DWT analysis can be performed by means of a two-channel subband coding scheme using a special class of filters called quadrature mirror filters as proposed by Mallat [38]. This is achieved by high-pass and low-pass filtering of the input signal by decomposing the signal into a coarse approximations and details as shown in Figure three. The coefficients for the approximations (*A*) and details (*D*) are calculated as follows:

$$A_{n,k} = 2^{-n/2} \sum_{i=1}^{j} x(i) \gamma(2^{-n} j - k), \quad (1)$$

$$D_{n,k} = 2^{-n/2} \sum_{i=1}^{j} x(i) \varphi(2^{-n} j - k), \quad (2)$$

where $\gamma$ is the scaling function, $\varphi$ the analysing mother wavelet, $j$ the number of discrete points of the input signal, $n$ and $k$ are the scaling index and the translation index, respectively [34]. The optimum decomposition level is determined here based on a minimum entropy algorithm as presented in [39].

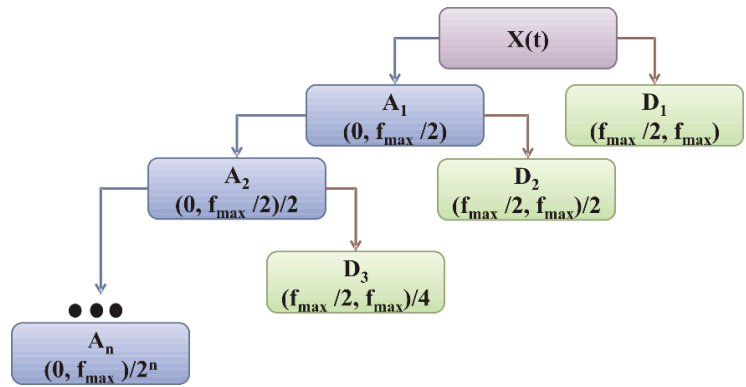

Figure 3. Decomposition Tree for the Discrete Wavelet Transform Algorithm.

Figure 4 and 5 show the signal reconstruction from different wavelet level decompositions of a structural dynamic response signal from a complex structure from the levels number one to nine. Further details about the structure can be found in [40]. It can be seen from Figure 4 that the highest level of decomposition corresponds to the denoising of the original analysed signal. Nevertheless, as one increases the number of level decompositions, the amount of coefficients is reduced. This is a great advantage since fewer coefficients are required in order to represent the relevant information contained in the recorded dynamic responses. Regarding the different signal reconstructions from the details decompositions shown in 5, it can be seen that waveforms emerge from the noise by increasing the level of decomposition.

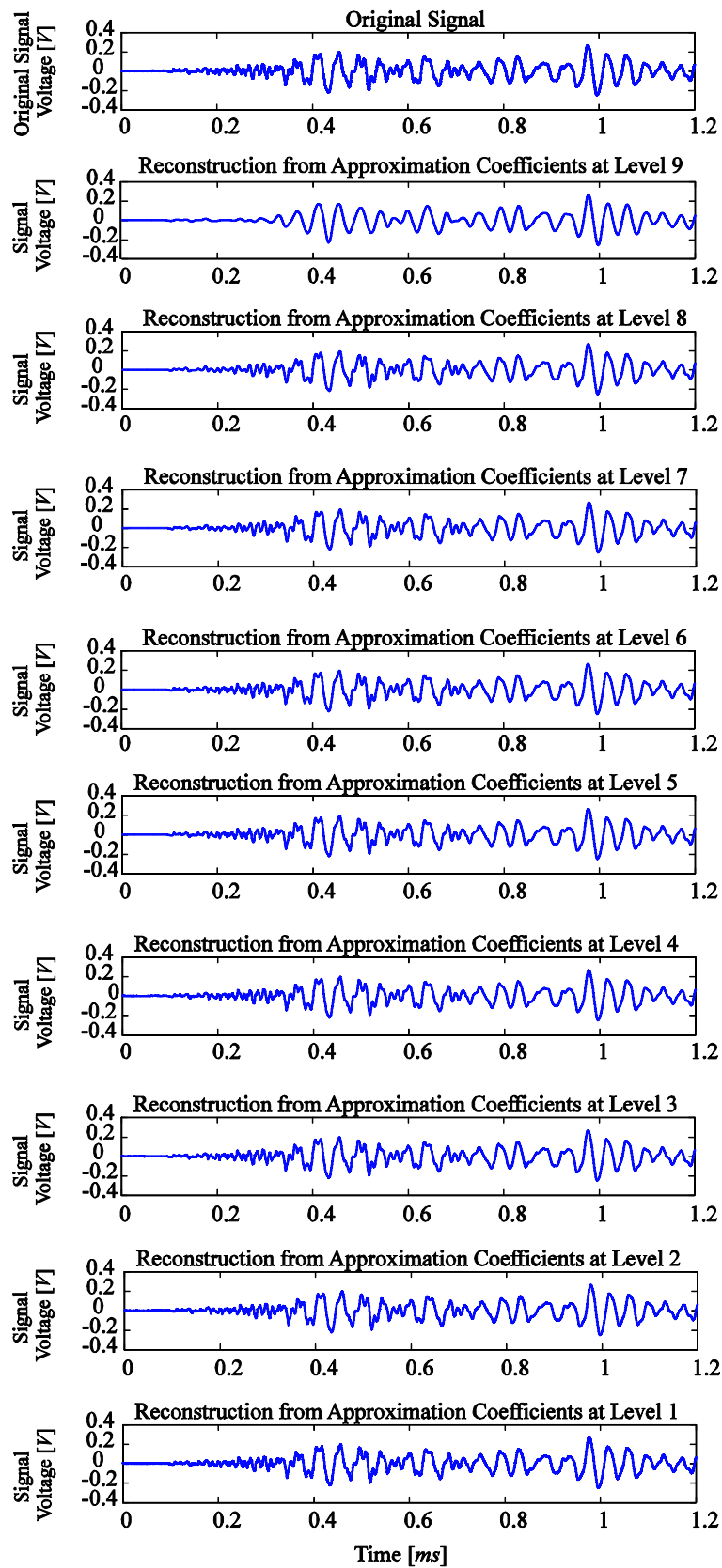

Figure 4. Wavelet decomposition and signal synthesis from approximation coefficients.

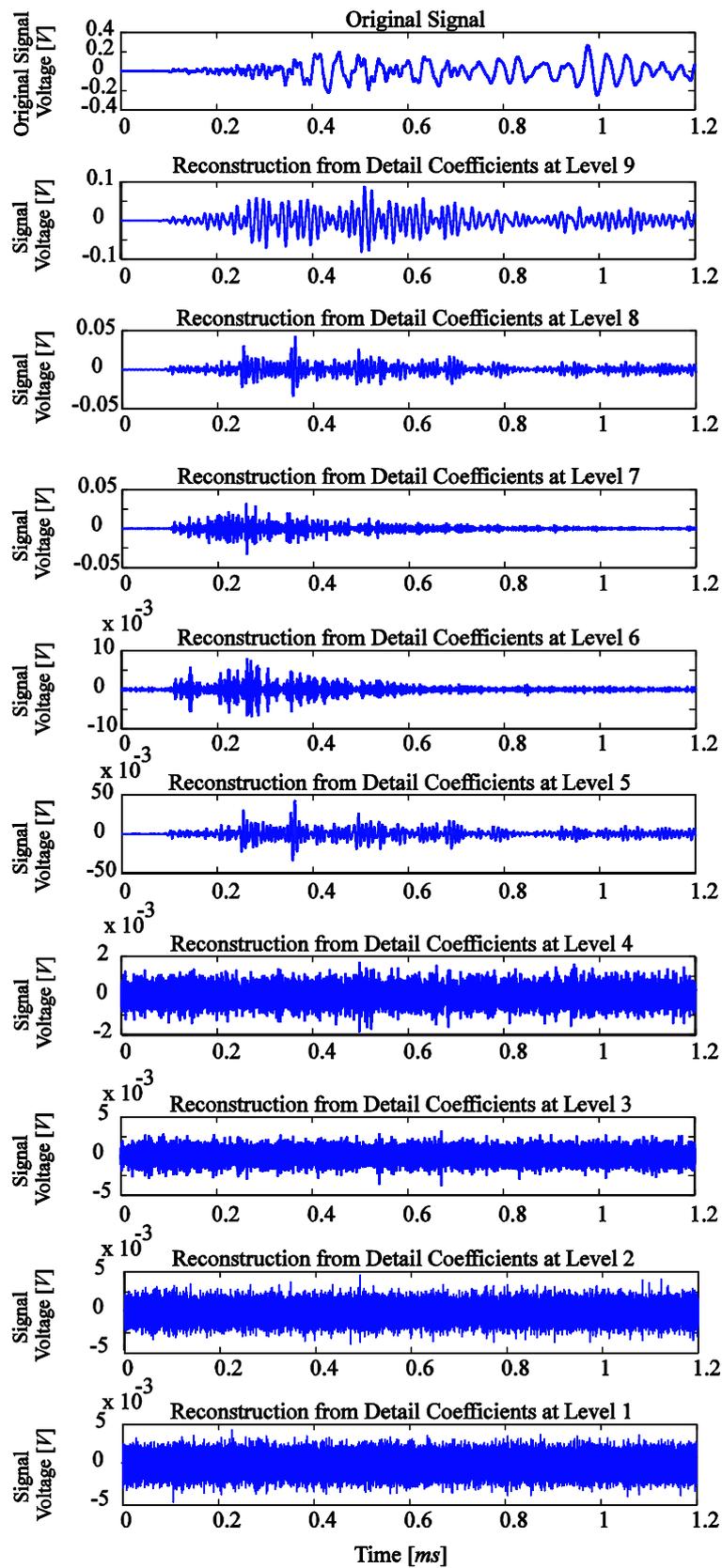

Figure 5. Wavelet decomposition and signal synthesis from detail coefficients.

*Data-Driven Modelling with Principal Component Analysis*

Principal Component Analysis (PCA) is well-known as a method for multivariate statistics [41]. Consider a matrix **X** with dimensions $n \times m$ containing the information from $m$ sensors and $n$ experiments. To apply PCA first, a normalization step of the **X** matrix should be considered [42]. Using this normalized matrix, the covariance matrix $\mathbf{C}_x$ can be then calculated as follows:

$$\mathbf{C}_x = \frac{1}{n-1}\mathbf{X}^T\mathbf{X}.$$

(3)

This is a square symmetric $m \times m$ matrix that measures the degree of linear relationship within the data set between all possible pairs of variables, i.e. sensors. The subspaces in PCA are defined by the eigenvectors and eigenvalues of the covariance matrix as follows:

$$\mathbf{C}_x\mathbf{P} = \mathbf{P}\mathbf{\Lambda},$$

(4)

where the eigenvectors of $\mathbf{C}_x$ are the columns of **P** and the eigenvalues are the diagonal terms of $\mathbf{\Lambda}$. The columns of matrix **P** are sorted with regard to the eigenvalues by descending order (called *Principal Components*). Selecting a reduced number of these principal components ($r < n$), the reduced transformation matrix $\mathbf{\Xi}$ could be defined as a baseline model for the pristine structure.

The data matrix **T** expresses the projection of the input data over the direction of the principal components $\mathbf{\Xi}$ (see Figure 6). This is done as follows:

$$\mathbf{T} = \mathbf{X}\mathbf{\Xi}.$$

(5)

The original data can be recovered according to $\mathbf{X} = \mathbf{T}\mathbf{P}^T$ in case retaining all principal components. In the reduced case, **X** cannot be fully recovered but **T** can be back projected into the original space to obtain a new data matrix:

$$\tilde{\mathbf{X}} = (\mathbf{X}\mathbf{\Xi})\mathbf{\Xi}^T.$$

(6)

In order to define the optimal number of principal components required for building the model, an analysis of the variances retained in the components could be performed. Principal components contributing less than a certain percentage to the total variance of the data set could be discarded as a criterion to select the number of required components.

There exist a number of statistical indices that can give information about the accuracy of the model and/or the adjustment of each experiment to the model. One very well-known index that is commonly used to this aim is the squared prediction error (SPE) statistic [43]. The SPE index (also called Q-statistic) measures the variability that breaks the normal process correlation indicating an abnormal situation. Denoting $\mathbf{e}_i$ as the *i-th* row of the matrix $\mathbf{E}$, the SPE for each experiment can be defined as follows:

$$\mathbf{SPE}_i = \mathbf{e}_i \mathbf{e}_i^T = \mathbf{x}_i (\mathbf{I} - \mathbf{\Xi}\mathbf{\Xi}^T) \mathbf{x}_i^T, \qquad (7)$$

where $x_i$ denotes the i-th row of the matrix X and I the identity matrix.

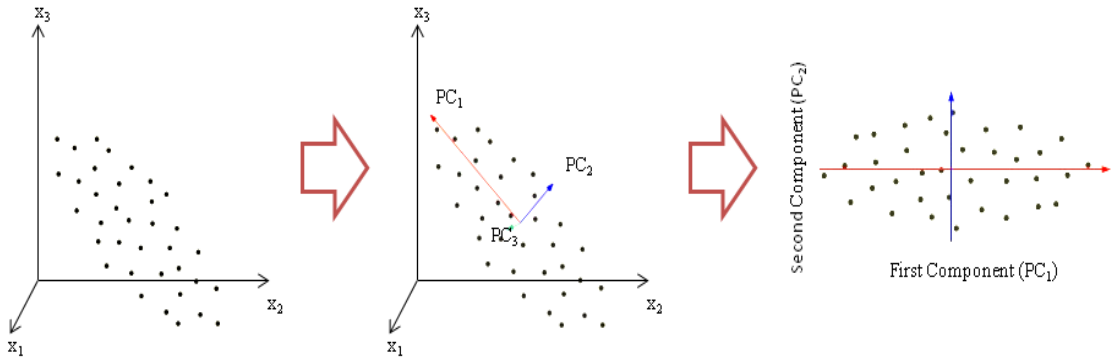

Figure 6. Projections of a set of three dimensional data points along the first two principal components (scores).

*Self Organizing Maps*

A self-organizing map (SOM) is a class of artificial neural network (ANN) with unsupervised learning, which purpose is to discover significant patterns in the input data without a target set. In its basic form, a SOM allows to convert the nonlinear relationships between high dimensional data into simple geometric relationships of their image points on a low dimensional display, usually, a regular two dimensional grid of nodes [35, 44].

A SOM compresses the information while preserving the most important topological and/or metric relationships of the primary data elements on the display, it may also be thought to produce some kind of abstractions. One of the most widely used SOM methodologies is the one developed by Kohonen. The goal of the Kohonen SOM is to transform an input pattern of arbitrary dimension in a bidimensional discrete map [45].

The main advantage of the SOM is its ability of permitting the grouping of input data into clusters. In order to achieve this goal, the SOM internally organizes the data based on features and their abstractions from input data.

The SOM is composed by elements called neurons. The neurons are located at the nodes of a bidimensional lattice and become tuned by several input patterns. Associated to each neuron is a weight vector, $m = (m_1, m_2, \dots, m_n)$, with same dimensions than the input data vectors ($n$) and, a position in the map space.

All the neurons are connected to adjacent neurons by a neighbourhood relation, which dictates the topology or structure of the map.

SOM applies competitive learning that means, neighbouring cells in a neural network compete in their activities by means of mutual lateral interactions, and develop adaptively into specific detectors of different signal patterns. When an observation is recognized, the activation of an output cell (competition layer) inhibits the activation of other neurons and reinforces itself. In simple words it is a "winner takes all" rule. [21]

The location of the tuned neurons tends to become ordered in such a way that a significant coordinate system for the features is created over the lattice. Hence, a SOM is characterized by the formation of a topographic map in which the spatial locations of the neurons, correspond to intrinsic features of the input patterns. [30]

The SOM is trained iteratively. In each step, one sample vector, $\bar{x}_m$, from the input data set is chosen randomly and the distance between this vector, and all the weight vectors of the SOM are calculated using different kind of distance measure. The neuron whose weight vector is closest to the input vector $\bar{x}_m$ is called Best-Matching Unit (BMU).

After finding the BMU, the weight factors of the SOM are updated so that, the BMU is moved closer to the input vector in the input space. The topological neighbours of the BMU are treated similarly. This process stretches the BMU and its topological neighbours towards the sample vector.

SOM uses the training process to organize the two dimensional maps consisting in the topological links between neurons connected by means of weights connections. The SOM can be used as a first phase of unsupervised classification or clustering.

*Density-based Simultaneous Two-Level - SOM*

A clustering process can be formally defined as the task of partitioning a set of objects into a collection of mutual disjoint subsets. Most clustering methods are able to performs an automatic detection of relevant sub-groups of clusters in unlabeled data sets, when no previous knowledge about the hidden structure of the data, is available. Usually, patterns into the same cluster should be similar to each other, while patterns in different clusters should not.

Cabanes *et al.* [46] proposed an efficient method of clustering based on the learning of a SOM. In the first phase the process, a standard SOM is used to compute a set of reference vector representing the local means of the data (weight vectors). Later, in a second phase, the obtained weight vectors are grouped in order to form the final partitioning. This approach is called a two-level clustering method.

The main idea of the two-level clustering technique based on SOM consists in combining the dimensionality reduction and learning capabilities of SOM with another clustering method applied to the reduced space, in order to produce a final set of clusters.

Perhaps, one of the most important task in clustering is to determine the number of clusters $K$. This task is also known as the model selection problem. If no previous knowledge about the data structure, there is no a simple way to estimate the number of clusters [36].

The methodology is based on learning at the same time the structure of the data and its segmentation by using both distance and density information. The algorithm assumes that a cluster is a dense region of objects surrounded by a region of low density. The main advantage of this methodology lies in the ability of the algorithm to determine automatically the number of clusters during the learning process. Then, no a priori hypothesis for the number of clusters is required. They called this particular methodology as "Local Density-based Simultaneous Two-Level Clustering" or DS2L-SOM.

As it was previously discussed, a SOM uses the training process to organize the two dimensional maps consisting in the topological links between neurons connected by means of weights connections. The mapping between the input space and the network space is then constructed in such way that two close observations in the input space would activate two close cells of the SOM. To achieve a topological mapping, the neighbours of a winner neuron can adjust their weight vectors towards the input data vector as well, but at a lesser degree, depending on how far away they are from the winner neuron. Usually a radial symmetric Gaussian neighbourhood function $K_{ij}$ is used for this purpose [36].

The cost function to be minimized with the DS2L-SOM is the distance between the input samples and the map reference vectors, weighted by a neighbourhood functions $K_{ij}$. This cost function is given by:

$$R(m) = \frac{1}{N} \sum_{n=1}^{N} \sum_{j=1}^{M} K_{j,u^*(\bar{x}_n)} \parallel m_j - \bar{x}_n \parallel^2,$$

(8)

where $M$ represents the number of neurons in the map, $N$ represents the number of learning samples, $u^*(\bar{x}_n)$ is the index of the neuron whose weight vector is the closest to the input vector $\bar{x}_n$ (the Best Matching Unit or BMU), and $K_{ij}$ is a positive symmetric kernel function. The relative importance of a neuron $i$ compared to the importance of a neuron $j$ is weighted by the value of this neighbourhood function, which can be expressed as:

$$K_{ij} = \frac{1}{\lambda(t)^e} e^{-\frac{d_{1(i,j)}^2}{\lambda^2(t)}},$$

(9)

where $\lambda(t)$ represents the "temperature function", which allows to modelling the topological neighbourhood extent. The temperature function is used in "Simulated Annealing Algorithms" (SAA) for finding the optimum configuration in an optimization problem.

In the DS2L-SOM algorithm each neighbourhood connection is associated with a real value $v$ which indicates the relevance of the connected neurons. This value is called "neighbourhood value" and is adapted during the learning process. The neighbourhood value of two weight vectors ($m_i$ and $m_j$), which is designated by $v_{i,j}$, is the number of data that are well represented by each one. $v_{i,j}$ can be expressed as the number of data having $i$ and $j$ as two first BMUs.
Then, for each data vector $\bar{x}_n$, the two closest weight vectors (BMUs) $u^*(\bar{x})$ and $u^{**}(\bar{x})$, are chosen:

$$u^*(\bar{x}) = argmin_i(Dist(m_i, \bar{x})),$$

(10)

$$u^{**}(\bar{x}) = argmin_{i \neq u^*(\bar{x})} \left( Dist(m_i, \bar{x}) \right).$$
(11)

For each data, both closest reference vectors are linked by a topological connection. The value of this connection will be increased, whereas the value of all other connections from the BMU will be reduced. At the end of the training, the set of interconnected reference vectors will constitute an artificial image of well separated clusters.

Each unit $i$ is also associated to an estimate of the local data density $D_i$, with the purpose of detect local density gradients which define the borders of neighbour clusters. Then, the "density modes" can be defined as a measure of the data density surrounding a weight vector which gives an idea about the amount of information present in an area of the input space.

For each data, $D_i$ is increased for all the units, as function of the Euclidean distance between the related weight vector $m_i$, and the data. The density estimation is given by:

$$D_i = \frac{1}{N} \sum_{n=1}^{N} \frac{e^{-\frac{Dist(m_i, \bar{x}_n)^2}{2\varrho^2}}}{\varrho\sqrt{2\pi}},$$
(12)

where $\varrho$ is a bandwidth parameter to be chosen by the user. If $\varrho$ is very big, all data will influence the density of all reference vectors, and close reference vectors will be associated to similar densities, resulting in worst accuracy of the estimate. On the other hand, if $\varrho$ is too small, a large portion of data (the most distant reference vectors) will not influence the density of the reference vectors, which induces a loss of information. It is recommended to use the average distance between a reference vector and its nearest neighbour as the value of $\varrho$ [47-48].

Besides the estimate the local density, the local variability is estimated for each weight vector. The variability $s$, is the mean distance between a weight vector $m$ and the $L$-th data $\bar{x}_m$, represented by $m$, in other words, it is the average distance between the weight vectors and the represented data:

$$s_i = \frac{1}{L} \sum_{i=1}^{L} Dist(m_i, \bar{x}_m).$$
(13)

Then, each weight vector of the SOM model is associated with a density and variability value, and each pair of weight vectors is associated with a neighbourhood value. After this, the data are clustered using the local density and the connectivity information in order to detect low-density boundaries between different clusters.

In Figure 7 an example of the sequence of the different stages of DS2L-SOM clustering algorithm is presented. The Figure 7(a) represents any arbitrary data set where, at naked eye, at least three different clusters seem to be present. Weight vectors linked together by a neighbourhood value $v > 0$ define well separated clusters. Each cluster is defined by a local maximum of density. This is useful to detect borders defined by large inter cluster distances (Figure 7(b)). The local density is used to detect cluster borders defined by low local density. Each cluster is defined by a local maximum of density (Figure 7(c)). Then, a "watershed" segmentation method (see below) is applied on density distribution of each well separated cluster in order to characterize sub-clusters (Figure

7(d)). For each pair of adjacent subgroups a density-dependant index is used to check if a low density area is a reliable indicator of the data structure, or whether, it should be regarded as a random fluctuation on the density (Figure 7(e)). The combined use of these two types of group definition can achieve very good results despite the low number of the prototypes in the map and is able to detect automatically the number of clusters (Figure 7(f)). [44]

According to [49], the "watershed" segmentation method was initially introduced by Lentuejoul in 1978. The method is an analogy of a natural watershed defined as an area where all the water that is under it, or drains off of it, goes into the same place. The original idea was to divide a digital image into sets using only the property of connectivity [46]. Finally, a verification is performed to identify if a low density regions is a reliable indicator of the data structure or if it is a random fluctuation of the density.

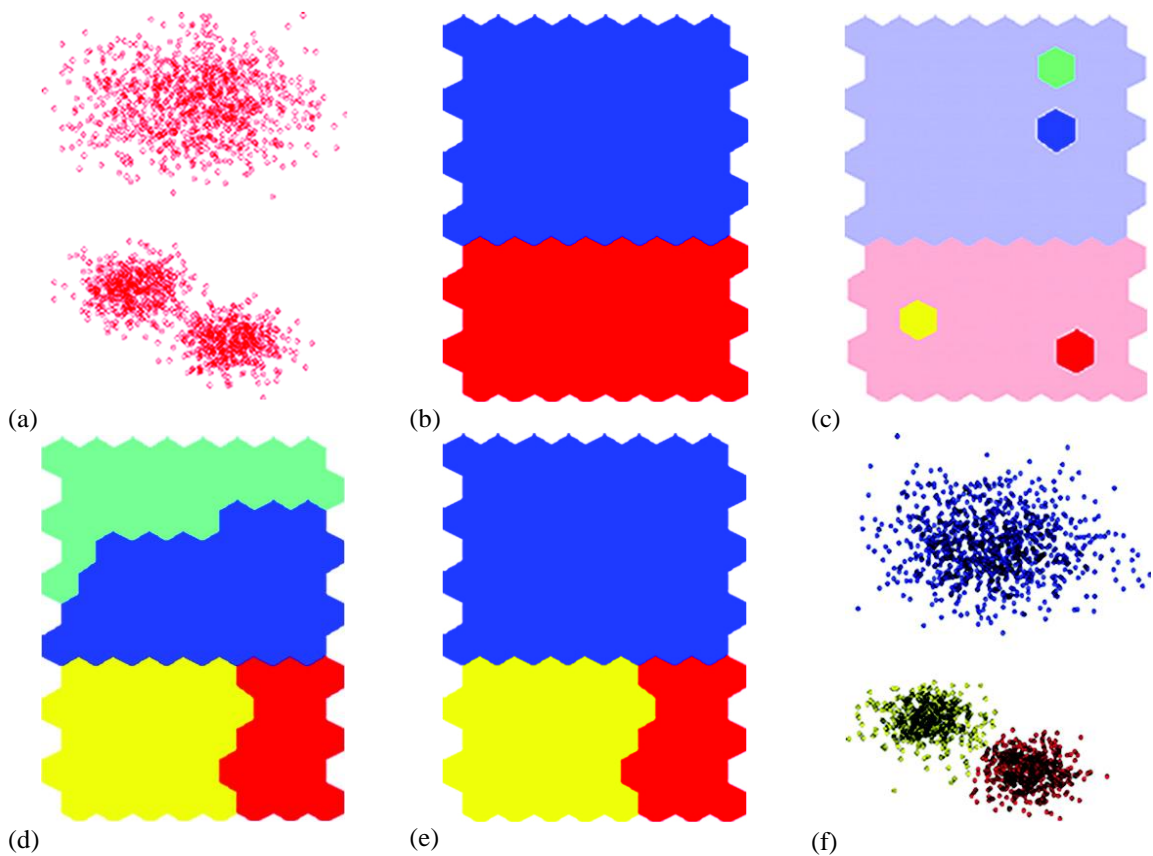

Figure 7. Example of sequence of the different stages of DS2L-SOM clustering algorithm: (a) Original data set, (b) Sets of connected prototypes, (c) Density modes detection, (d) Subgroups associated to each density mode, (e) Merging irrelevant groups: final clusters and (f) Data clustering from weight vectors clustering.

**PROPOSED METHODOLOGY FOR DAMAGE CLASSIFICATION**

The strategy proposed in this works differs from previously presented studies from the authors first by including the compensation of effects due to temperature changing conditions, and second by the use of the DWT together with the use of the DS2L-SOM, what provides an optimal scheme for sensor data fusion and automatic baseline selection within the framework of the proposed damage detection tasks (for previous works see [50]). The idea behind is to design a robust detector and try to tackle the problem related to the selection of an optimal baseline selection [51].

The health monitoring system presented in this work is based on a distributed array of permanently attached piezoelectric transducers where transducers are used in a pitch–catch configuration. As it was previously mentioned, an acousto-ultrasonics approach is used to collect all the waveform energy that is available. The structural dynamic responses recorded from an actuation step are stored and then pre-processed by the discrete wavelet transform (DWT), as a feature extraction technique, in order to calculate the approximation coefficients representing high scale components of the recorded signals. The optimum level of decomposition is calculated based on a minimum-entropy decomposition algorithm. The group of Daubechies wavelets ('db8') is evaluated as the analysing mother wavelet for this study [52]. This feature extraction procedure is repeated for all actuation steps and for all environmental conditions over a temperature range of these changing conditions in order to train a SOM in a subsequent step. It is good to stress here that the approximations coefficients are taken here for the analyses since they represent the interesting dynamics of the recorded waveforms and the detail coefficients will be considered as high-frequency noise.

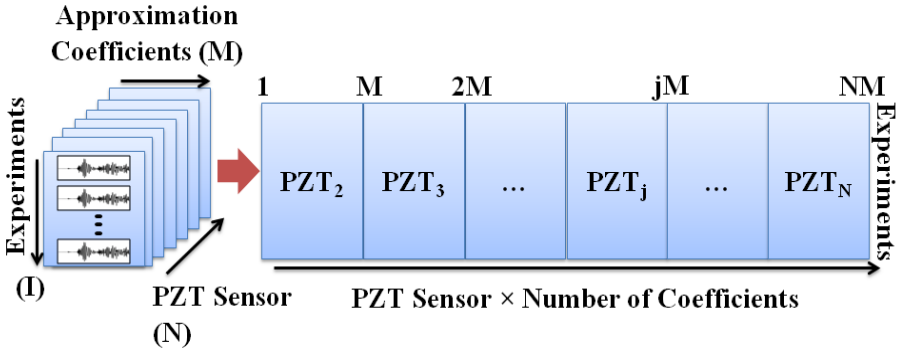

Figure 8. Unfolding procedure for sensor data fusion.

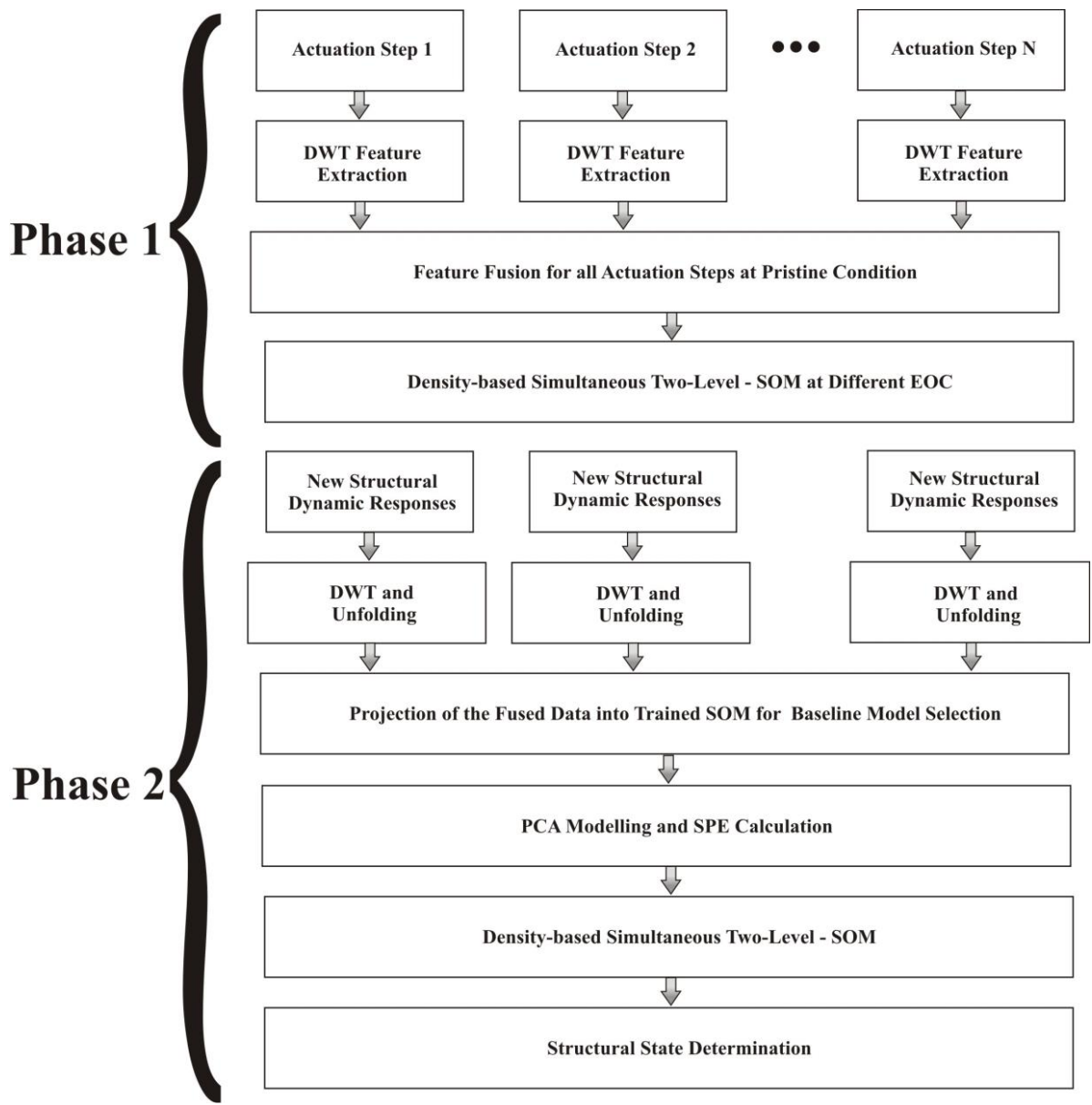

Figure 9. Flow diagram for the proposed methodology.

The DWT coefficients for each actuation step are calculated and fused following unfolding procedures (multiway) as it is done in multivariate statistical procedures for monitoring the progress of batch processes. This sensor data fusion procedure is depicted in Figure 8 within the first three steps contained in the first phase of the proposed strategy. Finally, the fused coefficients are used to train a SOM for each actuation step so that the different EOC can be clustered (with help of the DS2L-SOM algorithm) and serve as a basis for the optimal baseline selection which is required for the projection of new coming data so that they can be identified.

When new structural responses are available, from an unknown state, DWT approximation coefficients are extracted and projected into the respective trained self-organizing maps so that the optimal baseline can be selected (see phase two in Figure 8). If the data do not correlate with any cluster, then the new data are identified as coming from a damage state. Otherwise, a data-driven model based on Principal Component Analysis (PCA) is built with the features corresponding to this cluster [53]. To implement the PCA methodology, a normalisation of the collected data in each actuation step is performed first. Several studies of scaling for this kind of unfolded matrixes have been presented in the literature which include continuous scaling (CS), group scaling (GS) and auto-scaling (AS) [42]. In this paper, group scaling was used for normalization purposes. Finally, the new data is projected into the model (for each actuation step) and by retaining a certain number of principal components, squared prediction error (SPE) measures for all the actuation steps are calculated and used as input feature vectors for a self-organizing map (SOM) for the detection and identification tasks [54]. In order to calculate the number of clusters inside the data and provide a way for damage classification, a density-based simultaneous two-level clustering approach using SOM is evaluated [36]. Within this approach the structure of the data and its segmentation is learnt at the same time by using both distance and density information. The clustering algorithm assumes that a cluster is a dense region of objects surrounded by a region of low density. This approach is very effective when the clusters are irregular or intertwined, and when noise and outliers are present. The clustering algorithm divides automatically the dataset into a collection of subsets (clusters representing the pristine and damaged structural states) and the number of clusters is determined automatically during the learning process, i.e., no a priori hypothesis for the number of clusters is required [47]. The complete methodology is presented in Figure 9.

**EXPERIMENTAL SETUP AND RESULTS**

Experiments were performed for evaluating the performance of the presented methodology. The experiment used pairs of transducers operating in pitch-catch mode. The input signals to the PZT actuators were generated using the arbitrary signal generation capability of a combined signal generator and oscilloscope instrument manufactured by TiePie Engineering, Holland. The time histories were digitized at a sampling frequency of 50*MHz* and transferred to a portable PC for post-processing. The structure is a simplified aircraft composite skin panel made of carbon fibre reinforced plastic (CFRP) depicted in Figure 10.

The overall size of the plate is approximately 500×500×1.9*mm* and its weight is about 1.125*kg*. The stringers are 36*mm* high and 2.5*mm* thick. The plate and the stringers consist of 9 plies. Four piezoelectric transducers PIC-151 from PI Ceramics were installed on the surface of the plate with equidistant spacing. The piezoelectric transducers had a diameter of 10*mm* and a thickness of 0.5*mm*. They are separated by approximately 330*mm*. Damage conditions were simulated by placing magnets with different masses at the same position (see the red dot in Figure 4 to identify the position of the damage) on the structure as artificial damage in four increasing steps of 0.024*kg*. The damage was located with a distance of approximately 100*mm* from sensor number one. The aim of this form of artificial damage is to introduce reversible changes in the structures along the wave propagation paths. Temperature was varied from 35°C to 75°C in steps of 10°C for a total of five temperature levels. The excitation voltage signal is a 12*V* Hanning windowed cosine train signal with 5 cycles and carrier frequency of 50*kHz*. To determine the carrier central frequency for the actuation signal in each structure, a frequency sweep was performed and the spectral content of each signal was analysed. The carrier frequency was chosen to maximize the propagation efficiency. This type of excitation generates a dominant $A_0$ mode that is propagated along the structure allowing a better interaction of the guided wave with the simulated damage. A total of 720 signals per actuation step were recorded for every experiment at every temperature condition. A total of 120 thousand samples were taken per recoded signal what corresponds to recording time of approximately 2.4*ms* assuring that the whole signal was recorded. In a first step, all the collected baseline signals were processed by means of the DWT (a 8th level was found to be the optimum in all experiments), fused and presented to the DS2L-SOM algorithm in order to depict the changes introduced by the changing temperature.

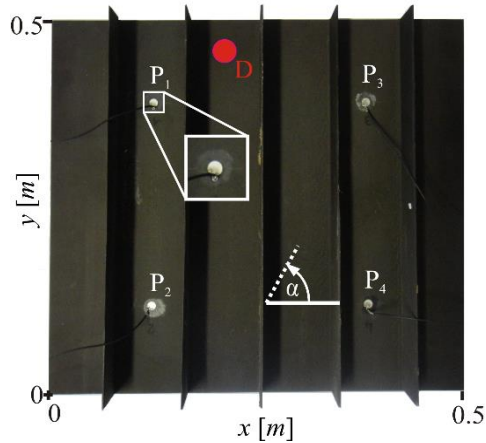

Figure 10. Experimental Setup: Stiffened Composite Panel.

As it can be seen from Figure 11(a) and (b), all the temperature levels were properly separated and distinguished with the help of the DS2L-SOM algorithm. Additionally, the U-Matrix, showing the average distance of a cell to its neighbouring cells (Figure 11(a)), allowed depicting the difference between the different formed clusters. This result is of special attention in baseline-based methods where the detection and characterization of damage is performed normally by means of metric indices by comparison of two dynamic response signatures what can lead to trigger a false alarm for damage detection just because of a change in environmental conditions.

In this work, each PCA model was generated using 70% of the whole data set collected from the undamaged structure. Signals from the remaining 30% are used for the validation of the models. A review of the variances retained in the components was performed in order to define the optimal number of components required for building the models from the pristine structural condition. This analysis is very important in order to ensure that enough variance is retained in the model that allows performing an optimal reduction. It is well known that if few principal components are selected, a poor model will be obtained leading to an incomplete representation of the process to be modelled. Nevertheless, if too many principal components are used, the model will be over-parameterized and will include noise. A threshold of 95% of retained cumulative variance was used in order to select the number of retained components [41].

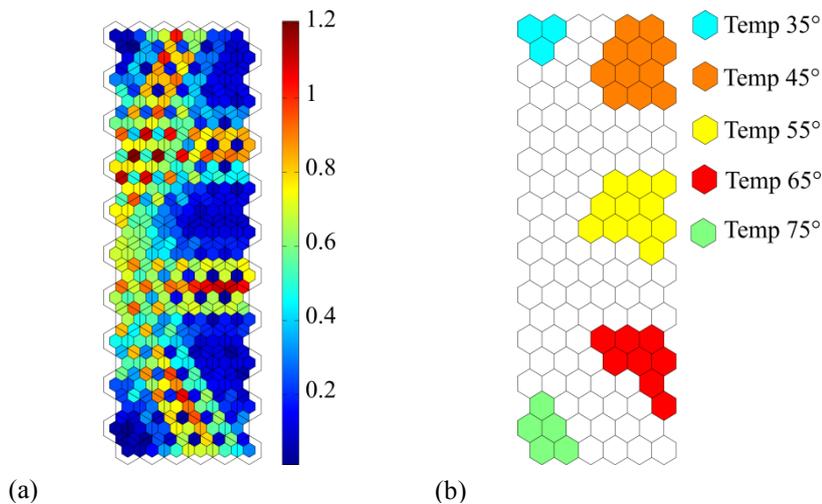

(a)　　　　(b)

Figure 11. All baselines presented to the DS2L_SOM algorithm: (a) U-Matrix and (b) DS2L-SOM cluster map.

Later on, four damage evolution steps were simulated using a local mass increase as it was defined before and at different temperature levels. From this point a second experiment was conducted. The idea of this experiment is to depict the effect of using a unique huge baseline model containing all possible temperature conditions. Without any loss of generality, all data belonging to the damages recorded at 35°C were used for the evaluation. Additionally, the baseline models were built with all the data coming from different temperature conditions, i.e. from 35°C to 75°C (without the optimal baseline selection step). A review of the variances retained in the components was performed in order to define the optimal number of components required for building the PCA models from the pristine structural condition. It was found that the first three hundred components included around 95% of the total variance into the reduced models for each actuation step, i.e. almost all the calculated components. As it is depicted in Figure 12(a) and (b), all the damaged states were not only separated from the pristine condition but also between them. Nevertheless, due to the big amount of data used for building the baseline models, a huge number of retained components (almost all) are required for obtaining an accurate approximate model. Another effect that can be seen from this figure is that the clusters are group around several neurons as a result of using almost all the recorded data into the model and hence, increasing the variability inside of it. However, even when the results depicted here are satisfactory since all damages are identified, the implementation of such a system in reality could be prohibited due to the big amount of computational power and storage capacity required. This applies especially for the case of remote monitoring units whose processing and storage characteristics are limited.

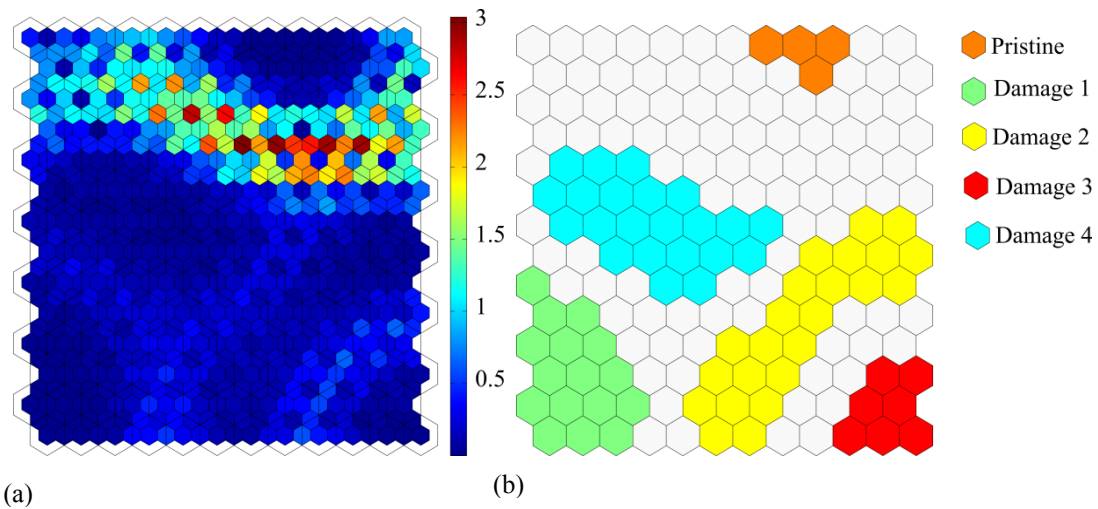

Figure 12. Damage detection with baseline models built with all the temperature range (35°C-75°C) for the local mass increase damage scenarios: (a) U-Matrix and (b) DS2L-SOM cluster map.

With the purpose of making the damage analyses more complicated, an additional experiment was performed. For this experiment, different damages were simulated on the surface of the plate at different positions and measurements were taken at different temperatures (35°C-75°C). After this, a general model was constructed with all the recorded data. The details for the damage locations are shown in Table 1. As it can be seen from Figure 13(a), the U-Matrix shows a clearly separation

(high boundary values) between the undamaged case and the simulated damaged scenarios. However, the boundaries between the different simulated damages are not so clearly defined, maybe explained by the fact of the great variability introduced by creating a model with all the recorded data. Additionally, it can be observed that some false cluttering results are present, i.e. data belonging to damage one are identified as undamaged. It can be seen as well how some damage cases are not properly identified and mixed (see Figure 13(b) – damage one, two and four). In this case again, almost all components are required in order to account for sufficient variance so that a robust model could be built.

| Damage No. | Location of damage w.r.t. PZTs No. | Coordinates of damage position on the CFRP plate | |
|---|---|---|---|
| | | *X position(mm)* | *Y position(mm)* |
| 1 | 1-2 | 125 | 250 |
| 2 | 2-4 | 225 | 125 |
| 3 | Around the centre of 1-2-3-4 | 295 | 225 |
| 4 | 3-4 | 375 | 250 |
| 5 | Beyond the line of 1-3 | 250 | 425 |

Table 1. Coordinates on specimen with respect to the damage numbers.

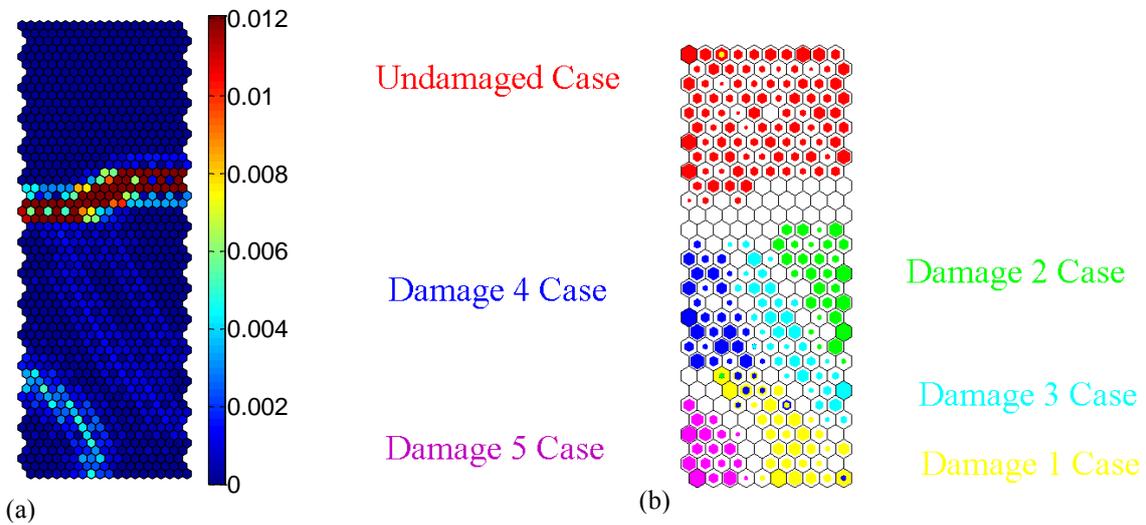

Figure 13. Damage detection with baseline models built with all the temperature range (35°C-75°C) and different simulated damages at different positions: (a) U-Matrix and (b) Evaluation of the DS2L-SOM algorithm.

In order to depict how this effect could be mitigated, new models were optimally selected and built according to the proposed methodology which is able to select baseline measurements lying in the range where the records belonging to a damage state were taken, i.e. around the optimal temperature. In this case, it was found that only the first thirty components included around 95% of the total variance into the reduced models for each actuation step. Figure 14(a) and (b) show, that here again, the proposed methodology was capable of separating all the damage cases from the pristine condition with the additional advantage of using a much more compact and reduced model in comparison to the previous example. In comparison to the previous example, the clusters are distributed more compactly around few neurons. This approach ensures not only a decrement in the computational cost but also provides robustness in the treatment of data by the reducing the variability into the model for the purpose of analyzing the state of the structure.

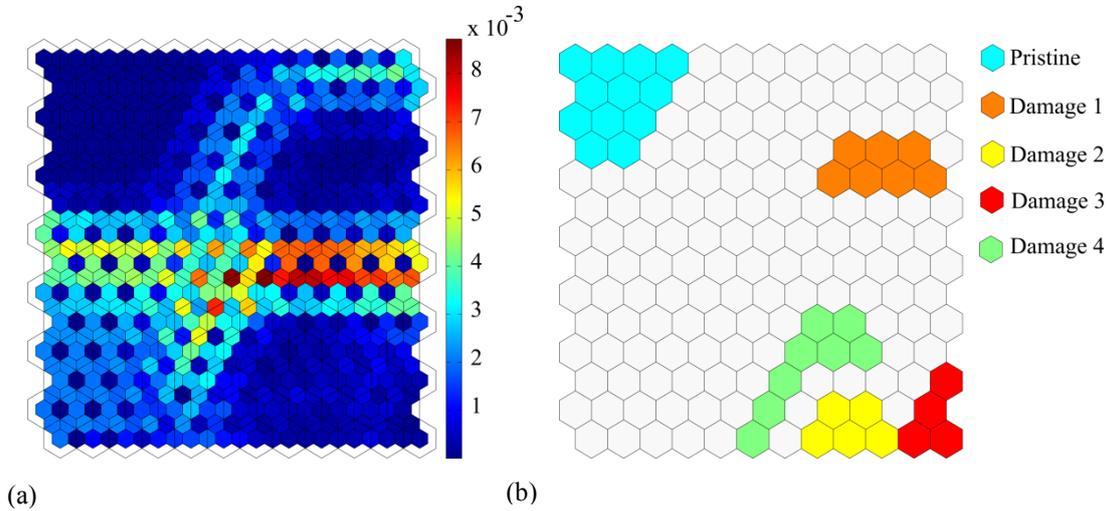

(a)  (b)

Figure 14. Damage detection with baseline models built with the proposed baseline selection methodology for the local mass increase damage scenarios: (a) U-Matrix and (b) DS2L-SOM cluster map.

One additional analysis, using with the damage descriptions described in Table 1 at different temperature levels (similar to the case evaluated in Figure 13), was performed in order to evaluate the proposed methodology. For this example, twenty five components were retained in order to build the baseline models at each temperature level whose input data were automatically clustered by the DS2L-SOM algorithm. Figure 15(a) shows the U-Matrix after running the complete processing algorithm. As it can be seen from this figure, all structural estates could be properly identified by well-defined boundaries. This is further corroborated by the cluster map depicted in Figure 15(b).

Receiver operating characteristics (ROC) curves were analysed to depict the advantages of using the presented methodology in comparison to standard the method based on PCA projections. The evaluation of PCA was motivated by the fact of its simplicity and low computational cost in comparison to other analysis techniques previously evaluated by the authors [55]. The area under the ROC curve (AUC) serves as comparison parameter, being a measurement of the accuracy of the classifier [56]. The value of AUC is always between zero and one. If the AUC is close to one, the classifier presents a very good diagnostic test. The AUC represents the probability that the classifier will evaluate randomly chosen positive instance higher than a randomly chosen negative instance. For calculating the ROC curves the code published by Giuseppe Cardillo was used [57]. As an evaluation, all the data collected during the experiments, i.e. all simulated damages and pristine condition at all evaluated temperature conditions were evaluated by using the proposed methodology. Additionally, these results were compared with the results produced by using the standard PCA as it can be found in literature.

Figure 16 shows the ROC study for the standard procedure based on PCA as presented by [58] and for the proposed methodology. As it can be seen, the AUC shows a higher value for the four actuation steps in the proposed methodology. A false positive ratio analysis was also performed and it was found that for the standard procedure the ratios were 16.4%, 21.9%, 15.4% and 26% for the actuation steps 1, 2 3 and 4, respectively. For the proposed method the ratios were 2.6%, 0.3%, 4.2% and 1.7% for the actuation steps 1, 2, 3 and 4, respectively. These results validates that a better discrimination may be possible if baseline signals at similar temperatures are used. It can be seen how the area under the ROC curve is increased for every actuation step. Another way to see that all actuation steps have a better performance is given by the fact that the ROC curves in Figure 16(b) are closer to the upper left corner compared to the curves depicted in Figure 16(a).

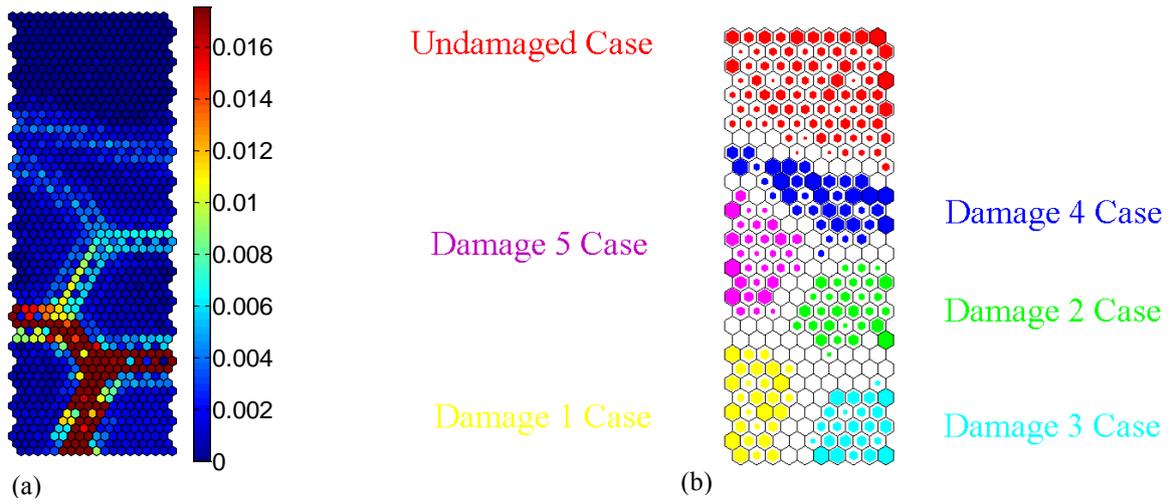

Figure 15. Damage detection with baseline models built with the proposed baseline selection methodology and evaluated with input data covering all the temperature range (35°C-75°C) and different simulated damages as in Table 1: (a) U-Matrix and (b) Evaluation of the DS2L-SOM algorithm.

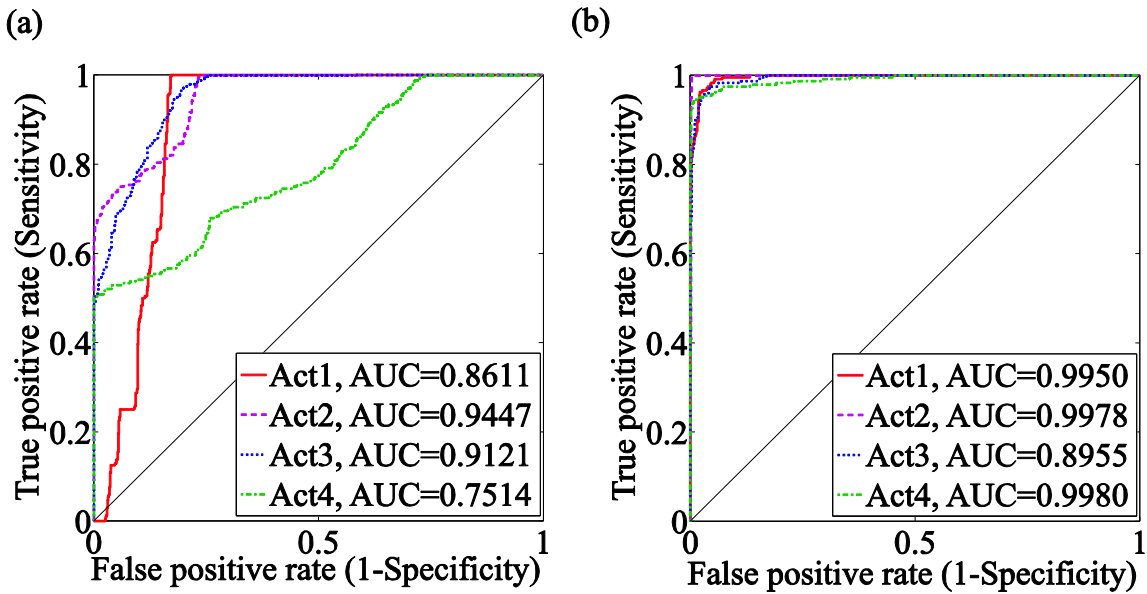

Figure 16. Receiver Operating Curves: (a) Results from the standard procedure and (b) Results from the proposed methodology.

CONCLUSIONS

The goal of this study was to illustrate and further develop a methodology to counterbalance the effects of environmental sources of variability on the performance of baseline data-driven models within the context of structural health monitoring systems and damage detection algorithms. The approach includes the combination of Discrete Wavelet Transform, Multi-Way Principal Component Analysis, Squared Prediction Error measures and Self-Organizing Maps. The evaluation of PCA was motivated by the fact of its simplicity and low computational cost in comparison to other analysis techniques previously evaluated by the authors. The coefficients extracted from the DWT allowed a reduction in the computational cost by decreasing the size of the unfolded matrices by each actuation step since the whole recorded time histories are not used. This step allowed lessening the difficulty of analysing directly the complex time traces by extracting relevant information and reducing the dimensionality of the problem. It was noticed that by including several temperature conditions at the same time in different models caused a decrease in the damage detection sensitivity since a model with more variability was built instead of specific models for each temperature condition, i.e. more scores, if not all of them, were required to build robust models. This effect motivated the necessity of developing the proposed classification and clustering methodology where an optimal baseline selection is accomplished. The automatic classification technique based on SOM and DS2L-SOM algorithms proved being effective in the detecting and classifying the different structural states under different temperature conditions. Future work will involve the improvement of the proposed methodology by including more advanced techniques for temperature compensation.

ACKNOWLEDGEMENT

The authors would like to express their gratitude to the Research School on Multi Modal Sensor Systems for Environmental Exploration (MOSES) and the Centre for Sensor Systems (ZESS), especially to Prof. Dr.-Ing. Otmar Loffeld, for sponsoring the research presented herein. The authors would like to thank Mr. Raghavendra Singh from the University of Siegen for his help during the experiments. Furthermore, the authors thank Prof. Dr.-Ing. C.-P. Fritzen from the University of Siegen for providing the access to the composite structure and his valuable suggestions.

REFERENCES

[1] Michaels, J.E., S.J. Lee, A.J. Croxford, and P.D. Wilcox, *Chirp excitation of ultrasonic guided waves.* Ultrasonics, 2013. **53**(1): pp. 265-270.
[2] Michaels, J.E., *Detection, localization and characterization of damage in plates with an in situ array of spatially distributed ultrasonic sensors.* Smart Materials and Structures, 2008. **17**(3).
[3] Flynn, E.B. and M.D. Todd, *Optimal Placement of Piezoelectric Actuators and Sensors for Detecting Damage in Plate Structures.* Journal of Intelligent Material Systems and Structures, 2010. **21**(1): pp. 265-274.
[4] Zhu, X. and P. Rizzo, *Guided waves for the health monitoring of sign support structures under varying environmental conditions.* Structural Control and Health Monitoring, 2011. **10**(1): pp. 1-17.
[5] Sohn, H., *Effects of environmental and operational variability on structural health monitoring.* Philosophical Transactions of the Royal Society A: Mathematical, Physical and Engineering Sciences, 2007. **365**(1851): pp. 539-560.

[6] Lu, Y. and J.E. Michaels, *Feature Extraction and Sensor Fusion for Ultrasonic Structural Health Monitoring Under Changing Environmental Conditions.* IEEE Sensors Journal, 2009. **9**(11): pp. 1462-1471.

[7] Fritzen, C.-P., G. Mengelkamp, and A. Guemes. *Elimination of temperature effects on damage detection within a smart structure concept.* In *Proceedings of the 4th International Workshop on Structural Health Monitoring*, Stanford, CA: DEStech Publications, Inc., 2003, pp. 1530-1538.

[8] Moll, J., P. Kraemer, and C.-P. Fritzen. *Compensation of Environmental Influences for Damage Detection using Classification Techniques.* In *Proc. 4th European Workshop on Structural Health Monitoring*, Krakow, Poland: DEStech, 2008.

[9] Buethe, I., P. Kraemer, and C.-P. Fritzen, *Applications of Self-Organizing Maps in Structural Health Monitoring.* Key Engineering Materials, 2012. **518**: pp. 37-46.

[10]. Torres Arredondo, M.A. and C.-P. Fritzen. *Ultrasonic guided wave dispersive characteristics in composite structures under variable temperature and operational conditions.* In *Proceedings of the 6th European Workshop in Structural Health Monitoring, EWSHM 2012*, Dresden, Germany, 2012, pp. 261-268.

[11] Lu, Y. and J.E. Michaels, *Discriminating damage from surface wetting via feature analysis for ultrasonic structural health monitoring systems.* Review of Progress in QNDE, 2008. **27**(1): pp. 1420-1427.

[12] Croxford, A.J., J. Moll, P.D. Wilcox, and J.E. Michaels, *Efficient temperature compensation strategies for guided wave structural health monitoring.* Ultrasonics, 2010. **50**(4-5): pp. 517-528.

[13] Kraemer, P., I. Buethe, and C.-P. Fritzen. *Damage detection under changing operational and environmental conditions using Self Organizing Maps.* In *Proceedings of SMART 11*, Saarbruecken, Germany, 2011.

[14] Torres Arredondo, M.-A., H. Jung, and C.-P. Fritzen. *Towards the development of predictive models for the system design and modal analysis of acoustic emission based technologies.* In *Proceedings of the 2nd International Workshop on Smart Diagnostics of Structures*, Krakow, Poland, 2011.

[15] Worden, K. and G. Manson, *The application of machine learning to structural health monitoring.* Philosophical Transactions of the Royal Society A: Mathematical, Physical and Engineering Sciences, 2007. **365**(1851): pp. 515-537.

[16] Farrar, C.R. and K. Worden, *An introduction to structural health monitoring.* Philosophical Transactions of the Royal Society A: Mathematical, Physical and Engineering Sciences, 2007. **365**(1851): pp. 303-315.

[17] Staszewski, W.J., *Advanced data pre-processing for damage identification based on pattern recognition.* International Journal of Systems Science, 2000. **31**(11): pp. 1381-1396.

[18] Moll, J., C. Heftrich, and C.-P. Fritzen, *Time-varying inverse filtering of narrowband ultrasonic signals.* Structural Health Monitoring, 2010. **9**(6).

[19] Vary, A. and K.J. Bowles, *An UItrasonic-Acoustic Technique for Nondestructive Evaluation of Fiber Composite Quality.* Polymer Engineering and Science, 1979. **19**(5): pp. 373-376.

[20] Finlayson, R.D., M. Friesel, M. Carlos, P. Cole, and J.C. Lenain, *Health Monitoring of Aerospace Structures with Acoustic Emission and Acousto-Ultrasonics.* Insight: Non-Destructive Testing and Condition Monitoring, 2001. **43**(3): pp. 155-158.

[21] Meyendorf, N., B. Frankenstein, and L. Schubert, *Structural health monitoring for aircraft, ground transportation vehicles, wind turbines and pipes - prognosis*, in *Emerging Technologies in Non-Destructive Testing V*, A.S. Paipetis, et al., Editors. London: Taylor and Francis Group, 2012.

[22] Torres Arredondo, M.-A., I. Buethe, D.-A. Tibaduiza, J. Rodellar, and C.-P. Fritzen, *Damage Detection and Classification in Pipework using Acousto-Ultrasonics and Non-*

| | |
|---|---|
| | *linear Data-Driven Modelling.* Journal of Civil Structural Health Monitoring, 2013. **3**(4): pp. 297-306. |
| [23] | Torres Arredondo, M.A., D.-A. Tibaduiza, L.E. Mujica, J. Rodellar, and C.-P. Fritzen, *Data-Driven Multivariate Algorithms for Damage Detection and Identification: Evaluation and Comparison.* International Journal of Structural Health Monitoring, 2014. **13**(1): pp. 5-18. |
| [24] | Torres Arredondo, M.A., *Acoustic Emission Testing and Acousto-Ultrasonics for Structural Health Monitoring.* PhD Thesis. Mechanical Engineering Department, University of Siegen, 2013. |
| [25] | Qiu, Y. and S. Backer, *Modeling for Suppression of Moisture/Temperature Induced Dimensional Changes in Fibrous Composite Structures.* Journal of Textile Institute, 2008. **86**(2): pp. 257-270. |
| [26] | Scalea, F.L.d. and S. Salamone, *Temperature effects in ultrasonic Lamb wave structural health monitoring systems.* Journal of the Acoustical Society of America, 2008. **124**(1): pp. 161-174. |
| [27] | Raghavan, A. and C.E.S. Cesnik, *Effects of Elevated Temperature on Guided-wave Structural Health Monitoring.* Journal of Intelligent Material Systems and Structures, 2008. **19**: pp. 1383-1398. |
| [28] | Kumar, B.G., R.P. Singh, and T. Nakamura, *Degradation of Carbon Fiber-reinforced Epoxy Composites by Ultraviolet Radiation and Condensation.* Journal of Composite Materials, 2002. **36**(24): pp. 2713-2733. |
| [29] | Jia, N. and V. Kagan, *Mechanical performance of polyamides with influence of moisture and temperature - accurate evaluation and better understanding*. 2003, BASF Corporation: New Jersey, USA. |
| [30] | Bank, L.C., T.R. Gentry, B.P. Thompson, and J.S. Russell, *A model specification for FRP composites for civil engineering structures.* Construction and Building Materials, 2003. **17**(6-7): pp. 405-437. |
| [31] | Wu, H.C., A. Yan, G. Fu, and R.F. Gibson. *Mechanics Based Durability Modeling of FRP Bridge Deck.* In *Proceedings of the Convention and Trade Show American Composites Manufacturers Association*, St. Louis, MO, USA, 2006, pp. 1-6. |
| [32] | Rose, J.L., *Ultrasonic Waves in Solid Media.* Cambridge: Cambridge University Press, 1999. |
| [33] | Moll, J., *Strukturdiagnose mit Ultraschallwellen durch Verwendung von piezoelektrischen Sensoren und Aktoren.* PhD Thesis. Mechanical Engineering Department, University of Siegen, 2011. |
| [34] | Mallat, S., *A Wavelet Tour of Signal Processing*. 2 ed. San Diego: Academic Press, 1997. |
| [35] | Kohonen, T., *Self Organizing Maps*: Springer, 2001. |
| [36] | Cabanes, G. and Y. Bennani, *Learning the Number of Clusters in Self Organizing Map*, in *Self-Organizing Maps*, G.K. Matsopoulos, Editor: InTech, 2010. |
| [37] | Worden, K., W.J. Staszewski, and J.J. Hensman, *Natural computing for mechanical systems research: A tutorial overview.* Mechanical Systems and Signal Processing, 2011. **25**(1): pp. 4-111. |
| [38] | Mallat, S.G., *A theory for multiresolution signal decomposition: the wavelet representation.* IEEE Transactions on Pattern Analysis and Machine Intelligence, 1989. **11**(7): pp. 674-693. |
| [39] | Coifman, R.R. and M.V. Wickerhauser, *Entropy-based algorithms for best basis selection.* IEEE Transactions on Information Theory, 1992. **38**(2): pp. 713-718. |
| [40] | Torres Arredondo, M.A. and C.-P. Fritzen. *Impact monitoring in smart structures based on gaussian processes.* In *Proceedings of the 4th International Symposium on NDT in Aerospace*, Augsburg, Germany (on CD-ROM), 2012. |
| [41] | Jollife, I.T., *Principal Component Analysis.* Springer Series in Statistics: Springer, 2002. |

[42] Westerhuis, J.A., T. Kourti, and J.F. MacGregor, *Comparing alternative approaches for multivariate statistical analysis of batch process data.* Journal of Chemometrics, 1999. **13**(3-4): pp. 397-413.
[43] Qin, S.J., *Statistical process monitoring: basics and beyond.* Journal of Chemometrics, 2003. **17**(8-9): pp. 480-502.
[44] Kaski, S., J. Nikklia, and T. Kohonen. *Methods for interpreting a self-organized map in data analysis*. In *Proceedings of the 6th European Symposium of Artificial Neural Networks (ESANN98)*, Brussels, Belgium, 1998, pp. 185-190.
[45] Reed, S., *Dimensionality reduction using linear and nonlinear transformation*, in *Encyclopedia of Structural Health Monitoring*: Wiley, 2009. pp. 757-770.
[46] Cabanes, G. and Y. Bennani, *Coupling clustering and visualization for knowledge discovery from data*, in *Proceedings of the International Joint Conference on Neural Networks IJCNN 11*. 2011.
[47] Cabanes, G., Y. Bennani, and D. Fresneau, *Enriched topological learning for cluster detection and visualization.* Neural Networks, 2012. **32**: pp. 186-195.
[48] Cabanes, G. and Y. Bennani. *A simultaneous two-level clustering algorithm for automatic model selection*. In *Proceedings of the Sixth International Conference on Machine Learning and Applications* 2007, pp. 316-321.
[49] Watson, A., *A new method of classification for landsat data using the "watershed" algorithm.* Pattern Recognition Letters, 1987. **6**: pp. 15-19.
[50] Tibaduiza, D.A., L.E. Mujica, and J. Rodellar, *Damage classification in structural health monitoring using principal component analysis and self-organizing maps.* Structural Control and Health Monitoring, 2012. **20**(10): pp. 1303-1316.
[51] Bishop, C.M., *Pattern Recognition and Machine Learning*. 1st ed. Information Science and Statistics. Singapore: Springer, 2007.
[52] Kalogiannakis, G. and D.v. Hemelrijck, *Classification of wavelet decomposed AE signals based on parameter-less self organized mappig.* International Journal of Materials and Product Technology, 2011. **41**(1-4): pp. 89-104.
[53] Tibaduiza, D.-A., M.-A. Torres Arredondo, L.E. Mujica, J. Rodellar, and C.-P. Fritzen, *A Study of Two Unsupervised Data Driven Statistical Methodologies for Detecting and Classifying Damages in Structural Health Monitoring.* Mechanical Systems and Signal Processing, 2013. **41**(1-2): pp. 467-484.
[54] Torres Arredondo, M.-A., D.-A. Tibaduiza, M. McGugan, H. Toftegaard, K.-K. Borum, L.E. Mujica, J. Rodellar, and C.-P. Fritzen, *Multivariate data-driven modelling and pattern recognition for damage detection and identification for acoustic emission and acousto-ultrasonics.* Journal of Smart Materials and Structures, 2013. **22**(10): pp. 1-21.
[55] Anaya, M., D.-A. Tibaduiza, M.-A. Torres Arredondo, F. Pozo, M. Ruiz, L.E. Mujica, J. Rodellar, and C.-P. Fritzen, *Data-driven methodology for damage detection and classification under temperature changes.* Submitted to Journal of Computers and Structures, 2013.
[56] Mujica, L.E., M. Ruiz, F. Pozo, J. Rodellar, and A. Güemes, *Structural damage detection indicator based on principal component analysis (PCA) and statistical hypothesis testing.* Submitted to Structural Control and Health Monitoring, 2013.
[57] Cardillo, G. (2013) *ROC Curve for Matlab*.
[58] Mujica, L.E., J. Rodellar, A. Fernandez, and A. Guemes, *Q-statistic and T2-statistic PCA-based measures for damage assessment in structures.* Structural Health Monitoring, 2010. **10**(5): pp. 539-553.